\pdfoutput=1

\documentclass[aps,twocolumn,amsmath,amssymb,preprintnumbers,floatfix,longbibliography]{revtex4-1}

\usepackage[utf8]{inputenc}
\usepackage{newtxtext}
\usepackage[upint]{newtxmath}
\usepackage{microtype}
\usepackage{textcomp}
\usepackage{eucal}
\usepackage{bm}
\usepackage{siunitx}
\usepackage{comment}

\usepackage{enumerate}
\usepackage{amsfonts}
\usepackage{amsmath}
\usepackage{amssymb}
\usepackage{color}
\usepackage{soul}

\usepackage{graphicx}

\usepackage[colorlinks,allcolors=blue]{hyperref}
\usepackage[capitalize]{cleveref}

\definecolor{DarkRed}{rgb}{0.65,0,0}%
\definecolor{Green}{rgb}{0,0.3,0.3}
\definecolor{Purple}{rgb}{0.3,0,0.65}
\definecolor{Red}{rgb}{1,0,0}
\definecolor{Blue}{rgb}{0,0,0.85}

\newcommand{\Imag}{{\Im\mathrm{m}}}   
\newcommand{\ve}[1]{\boldsymbol{#1}}

\newcommand{\veck}{\ve{k}}

\newcommand{\vek}{\ve{k}}

\newcommand{\eg}{\textit{e.g. }}

\newcommand{\be}{\begin{equation}}
\newcommand{\ee}{\end{equation}}

\newcommand{\veq}{\ve{q}}

\newcommand{\veQu}{\ve{Q}^{\uparrow}}
\newcommand{\veQd}{\ve{Q}^{\downarrow}}
\newcommand{\veQs}{\ve{Q}^{\sigma}}

\newcommand{\prlsection}[1]{\textit{#1}.\kern0.05em---\kern0.05em\ignorespaces}

\usepackage{stackengine}
\stackMath

\begin{document}
\title{Magnon spin current induced by triplet Cooper pair supercurrents}
\author{Lina G. Johnsen}
\email[Corresponding author: ]{lina.g.johnsen@ntnu.no}
\affiliation{Center for Quantum Spintronics, Department of Physics, Norwegian \\ University of Science and Technology, NO-7491 Trondheim, Norway}
\author{Haakon T. Simensen}
\affiliation{Center for Quantum Spintronics, Department of Physics, Norwegian \\ University of Science and Technology, NO-7491 Trondheim, Norway}
\author{Arne Brataas}
\affiliation{Center for Quantum Spintronics, Department of Physics, Norwegian \\ University of Science and Technology, NO-7491 Trondheim, Norway}
\author{Jacob Linder}
\affiliation{Center for Quantum Spintronics, Department of Physics, Norwegian \\ University of Science and Technology, NO-7491 Trondheim, Norway}
\date{\today}

\begin{abstract}
At the interface between a ferromagnetic insulator and a superconductor there is a coupling between the spins of the two materials. 
We show that when a supercurrent carried by triplet Cooper pairs flows through the superconductor, the coupling induces a magnon spin current in the adjacent ferromagnetic insulator. 
The effect is dominated by Cooper pairs polarized in the same direction as the ferromagnetic insulator, so that charge and spin supercurrents produce similar results.
Our findings demonstrate a way of converting Cooper pair supercurrents to magnon spin currents.
\end{abstract}
\maketitle


\prlsection{Introduction} Ferromagnetic insulators (FI) are of high relevance for spin transport applications due to their ability to carry pure spin currents over long distances \cite{cornelissen_np_15}. The study of how these spin currents can be converted into conventional electron-based charge and spin currents, and vice versa, has therefore attracted much attention over a number of years \cite{chumak_np_15,han_nm_20,brataas_pr_20}. 

Early research on the topic showed that spin-polarized electron currents can excite the magnetization of a metallic ferromagnet via a spin-transfer torque \cite{slonczewski_jmmm_96,berger_prb_96,tsoi_prl_98,manchon_rmp_19}, thus causing propagating spin waves \cite{demidov_nm_10,madami_nn_11,wang_prl_11}.
The spin-polarized electron current can be obtained by sending a charge current through a homogeneous magnetic region \cite{tsoi_prl_98}, or through the spin-dependent scattering of a charge current via the spin-Hall effect \cite{dyakonov_pla_71,ando_prl_08,kajiwara_n_10,demidov_apl_11}.
In this way, one may convert charge currents into magnon spin currents.
Conversely, precession of the magnetization of a FI cause spin-pumping where spin is injected into an adjacent normal-metal \cite{tserkovnyak_prl_02,simanek_prb_03,woltersdorf_prl_05,costache_prl_06,tserkovnyak_rmp_05}. The injected spin current can further be transformed into a charge current through the inverse spin-Hall effect \cite{saitoh_apl_06,sandweg_prl_11,czeschka_prl_11,castel_apl_11,hahn_prb_13,weiler_prl_13}.

The precession of the magnetization of a FI can similarly cause injection of quasi-particle spin currents into a conventional $s$-wave superconductor (S). Early reports showed a decrease in the spin injection causing a reduced Gilbert damping \cite{bell_prl_08}. This follows from the fact that the appearance of an energy gap below the superconducting critical temperature makes spin injection and transport in the form of quasi-particles possible only at energies above the gap edge \cite{hyunsoo_nm_10}. Contrary to these observations \cite{bell_prl_08,jeon_pra_18}, other works found a coherence peak in the Gilbert damping just below the superconducting critical temperature \cite{inoue_prb_17,yao_prb_18}. Theoretical works have followed the experimental advances \cite{kato_prb_19,silaev_prb_20,silaev_prb_20b,ahari_prb_21,ominato_arxiv_21} considering \eg non-colinear magnetizations \cite{simensen_prb_21}, and the effects of a spin-splitting field \cite{olajarvi_prb_20,heikkila_pss_19}. The latter has been shown experimentally to cause a giant enhancement of the spin transport across the interface \cite{jeon_ACSnano_20}.
The converse effect, a conversion from quasi-particle to magnon spin currents has been predicted when the population of one of the quasi-particle spin species is higher \cite{vargas_jmmm_20}.

Superconductors can however also carry charge and spin supercurrents \cite{eschrig_rpp_15}. In conventional singlet superconductors, the Cooper pairs carry zero net spin and are easily destroyed by magnetic fields. Supercurrents carried by spin-polarized triplet Cooper pairs are on the other hand robust to the spin dependent pair-breaking effect of magnetic fields and allow for dissipationless spin transport \cite{bergeret_rmp_05,buzdin_rmp_05,linder_np_15}. Moreover, they have attracted a lot of attention in the light of recent spin-pumping experiments where both spin-polarized supercurrents \cite{jeon_np_18,jeon_prb_19,jeon_pra_19,jeon_prb_19b} and quasi-particle spin currents \cite{muller_prl_21} have been pointed out as potential explanations for the observed effects.
Motivated by these advances, we pose the following fundamental question: Is it possible for a triplet Cooper pair supercurrent to induce a magnon spin current?

\begin{figure}[b]
    \centering
    \includegraphics[width=\columnwidth]{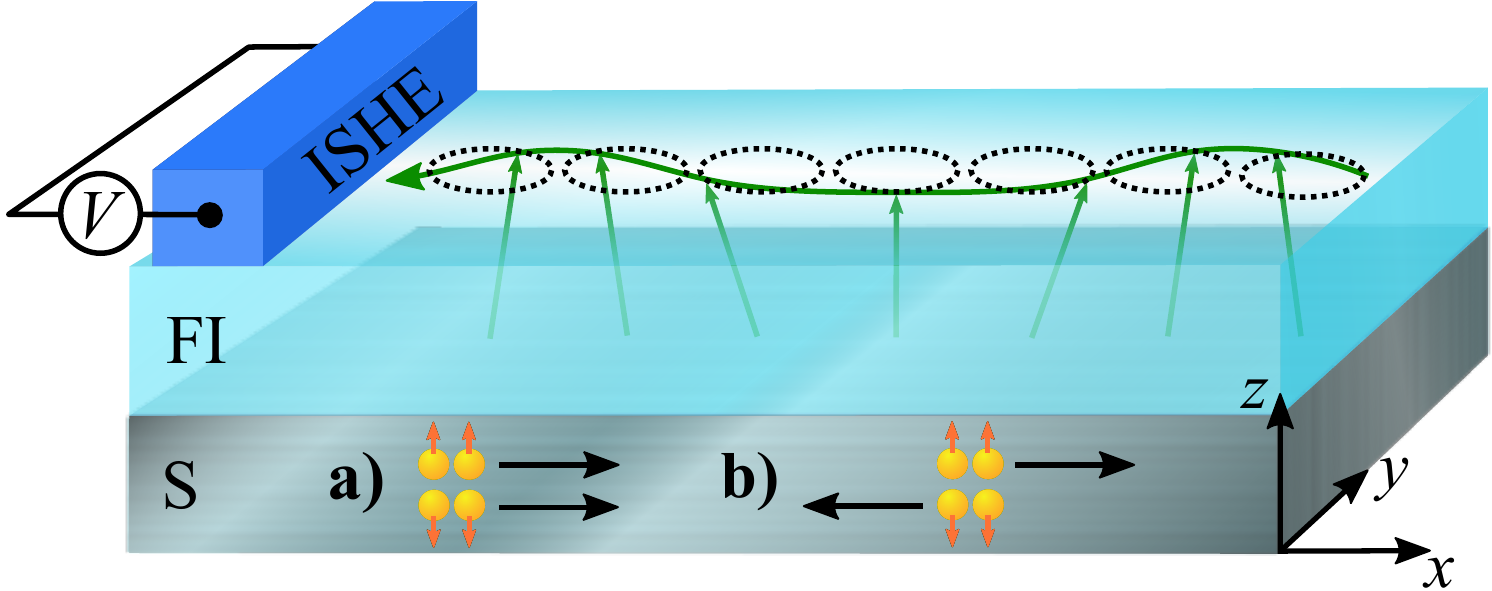}
    \caption{A superconductor (S) can carry: a) a charge supercurrent consisting of an equal number of spin-up and spin-down triplet Cooper pairs traveling in the same direction, or b) a spin supercurrent where the spin-up and spin-down triplets travel in opposite directions. Because of a coupling between the spins in the superconductor and in an adjacent ferromagnetic insulator (FI), these supercurrents induce a magnon spin current in the FI (green). The induced spin current can be measured through a normal-metal contact via the inverse spin-Hall effect (ISHE). In order to couple the currents, we only require that the Cooper pairs must carry a net spin along the easy axis of the FI which is fixed along the $z$ axis. The orientation of the interface is not restricted by our model, so that the Cooper pair-induced magnon spin current also occurs if the FI is instead polarized in-plane, in that case with an interface in the $xz$ plane.}
    \label{fig:system}
\end{figure}

To answer this question, we study the coupling between the spins in a ferromagnetic insulator and a $p$-wave superconductor carrying triplet Cooper pair supercurrents. 
We find that to the first order in perturbation theory, a supercurrent only causes a renormalization of the magnon gap. To the second order, both charge and spin supercurrents induce 
a magnon spin current in the adjacent FI. This effect is dominated by Cooper pairs polarized parallel to the magnetization.
The magnon spin current appears due to a symmetry breaking in the magnon energy spectrum with respect to momentum inversion. 
The asymmetry originates from the coupling to the superconductor, which due to the presence of the supercurrent has an energy spectrum that is tilted in momentum space.
We propose that the supercurrent-induced magnon spin current can be measured in the S/FI structure shown in Fig.~\ref{fig:system}.
However, as we will discuss further, an intrinsic triplet superconductor is not necessarily required, since such superconductivity can be established even in non-superconducting materials via the proximity effect \cite{bergeret_rmp_05,buzdin_rmp_05,eschrig_rpp_15,linder_np_15}.
In the following, we present our model for describing this structure.

\prlsection{Microscopic model}
To describe the S/FI structure, we need a Hamiltonian $H=H_{\text{S}}+H_{\text{FI}}+H_{\text{c}}$ describing the superconductor, the ferromagnetic insulator, and the coupling between these, respectively.
We describe this system as two separate translationally invariant two-dimensional (2D) layers on top of each other, where coupling between the layers exists for lattice sites corresponding to the same location in the plane. This allows us to describe our Hamiltonian in momentum space. Details on how the Hamiltonian can be derived from a lattice model is described in the Supplemental Material (SM) \cite{SM}.

We describe the current-carrying state of the superconductor by the mean-field Hamiltonian
\begin{align}
H_{\text{S}}&=\sum_{\veck,\sigma}\epsilon_{\veck}c_{\veck,\sigma}^{\dagger}c_{\veck,\sigma}\notag\\
&\phantom{=}-\frac{1}{2}\sum_{\ve{k},\sigma}[\Delta_{\vek,\sigma}^{\ve{\delta}}c_{\ve{k}+\ve{Q}^{\sigma},\sigma}^{\dagger}c_{-\ve{k}+\ve{Q}^{\sigma},\sigma}^{\dagger}+\text{h.c.}],\\
\epsilon_{\vek}&=-\mu-2t\sum_{\ve{\delta}}\cos(\vek\cdot\ve{\delta} ) ,
\hspace{0.5em}\Delta_{\vek,\sigma}^{\ve{\delta}}=f_{\sigma}^{\ve{\delta}} \sin(\vek\cdot\ve{{\delta}})
\end{align}
corresponding to $p_{x(y)}$-wave superconductivity for $\Delta_{\vek,\sigma}^{\ve{x}(\ve{y})}$. The first term in $H_{\text{S}}$ sets the chemical potential $\mu$ and describes hopping between nearest neighbor sites, where~$t$ is the hopping integral. 
For layers with a square lattice structure, the vectors $\ve{\delta}$ from a given lattice site to all nearest neighbors is a set of two perpendicular vectors. We can also describe a one-dimensional (1D) chain by eliminating one of the vectors in the set. 
The second term gives rise to $p$-wave superconducting triplet pairing between electrons of equal spin. The Cooper pairs have a center-of-mass momentum of $2\veQs$ so that we have a supercurrent of spin-$\sigma$ Cooper pairs inside the superconductor. The supercurrent is enforced by an externally applied homogeneous phase gradient \cite{SM,takashima_prb_17,lin_sc_14}, which can be obtained experimentally via electric contacts at each end of the superconductor. The applied current is assumed to be smaller than the critical supercurrent of the system. The strength of the superconducting order parameter $\Delta_{\vek,\sigma}^{\ve{\delta}}$ is given by the real parameter $f_{\sigma}^{\ve{\delta}}$. The operators $c^{(\dagger)}_{\vek,\sigma}$ annihilate (create) an electron with momentum $\vek$ and spin $\sigma$.

The Hamiltonian describing the ferromagnetic insulator is obtained from a Heisenberg model with uniaxial anisotropy favoring parallel spins aligned along the $z$ axis. By performing a Holstein-Primakoff (HP) transformation to the second order in the bosonic operators, we arrive at
\begin{align}
H_{\text{FI}}&=\sum_{\veq}\omega_{\veq}a_{\veq}^{\dagger}a_{\veq},
\hspace{0.5em}\omega_{\veq}=4S\Big\{K + J\sum_{\ve{\delta}}[1-\cos(\veq\cdot\ve{\delta} )] \Big\}.
\end{align}
The strength of the coupling between spins on neighboring lattice sites is $J>0$, and $S$ is the magnitude of the spins. The anisotropy favoring alignment along the $z$ axis gives rise to a magnon gap whose magnitude is determined by $K>0$. We include the magnon gap to make contact with an experimentally realistic setting, and also to satisfy the condition that the coupling strength must be small compared to the original magnon energies for all momenta. The operators $a^{(\dagger)}_{\veq}$ annihilate (create) a magnon of momentum $\veq$. Since the FI is polarized along the $z$ axis, all magnons have spin along $-\ve{z}$.

We now consider the coupling between the superconductor and the ferromagnetic insulator. By performing a HP transformation to the second order in the bosonic operators on the coupling $H_{\text{c}}=-\sum_{\ve{i}}\Lambda \ve{s}_{\ve{i}}\cdot \ve{S}_{\ve{i}}$ between the electron spins $\ve{s}_{\ve{i}}$ in the superconductor and the spins $\ve{S}_{\ve{i}}$ in the ferromagnetic insulator, we find that  the coupling $H_{\text{c}}=H_{\text{c}}^{2c}+H_{\text{c}}^{2c1a}+H_{\text{c}}^{2c2a}$ consists of three terms. These are given by
\begin{align}
    H_{\text{c}}^{2c}&=-\lambda\sqrt{\frac{NS}{2}}\sum_{\vek,\sigma}\sigma c_{\vek,\sigma}^{\dagger}c_{\vek,\sigma}\label{eq:Hc_2c}\\
    H_{\text{c}}^{2c1a}&=-\lambda\sum_{\vek,\veq}(c_{\vek+\veq,\uparrow}^{\dagger}c_{\vek,\downarrow}a_{-\veq}^{\dagger}+c_{\vek+\veq,\downarrow}^{\dagger}c_{\vek,\uparrow}a_{\veq}\label{eq:Hc_2c1a})\\
    H_{\text{c}}^{2c2a}&=\frac{\lambda}{\sqrt{2NS}}\sum_{\vek,\veq,\veq',\sigma}\sigma c_{\vek+\veq,\sigma}^{\dagger}c_{\vek-\veq',\sigma}a_{-\veq}^{\dagger}a_{\veq'}.\label{eq:Hc_2c2a}
\end{align}
Above, $N$ is the total number of lattice sites, and $\lambda=\Lambda\sqrt{S/2N}>0$ is the coupling strength, where we have absorbed a factor into the coupling strength for simplicity of notation.
The first term, $H_{\text{c}}^{2c}$, describes a spin-splitting of the fermionic energy spectrum. We absorb this term into $H_{\text{S}}$ by letting $\epsilon_{\vek}\to\epsilon_{\vek,\sigma}=\epsilon_{\vek}-\sigma\lambda\sqrt{NS/2}$. 
In the absence of magnons, $H_{\text{S}}+H_{\text{c}}^{2c}$ reproduces the familiar energy spectrum of a triplet superconductor in a spin-splitting field.
The second term, $H_{\text{c}}^{2c1a}$, transfers spin between fermion and boson operators and will turn out to be significant for inducing a magnon spin current. 
The third term, $H_{\text{c}}^{2c2a}$, only gives a constant shift in the magnon energy spectrum to the first order in perturbation theory and is projected out to the second order.

Before we start analyzing how the coupling affects the magnonic part of the Hamiltonian, we first write the superconducting part in the diagonal form
\begin{align}
    H_{\text{S}}=&\sum_{\vek,\sigma}E_{\vek-\veQs,\sigma}\gamma_{\vek,\sigma}^{\dagger}\gamma_{\vek,\sigma},\\
    E_{\vek,\sigma}=&\phantom{+}\frac{1}{2}(\epsilon_{\vek+\ve{Q}^{\sigma},\sigma}-\epsilon_{-\vek+\ve{Q}^{\sigma},\sigma})\notag\\
    &+\sqrt{\left[\frac{1}{2}(\epsilon_{\vek+\ve{Q}^{\sigma},\sigma}+\epsilon_{-\vek+\ve{Q}^{\sigma},\sigma})\right]^2 + \big|\Delta_{\vek}\big|^2  }.
    \label{eq:E}
\end{align}
We have assumed the magnitude of the superconducting order parameter $|\Delta_{\vek}|$ to be spin-independent.
The old operators are related to the new ones by
\begin{align}
    c_{\vek+\ve{Q}^{\sigma},\sigma}&=u_{\vek,\sigma}\gamma_{\vek+\ve{Q}^{\sigma},\sigma}+ \text{sgn}(\Delta_{\vek}^{\sigma})v_{\vek,\sigma}\gamma_{-\vek+\ve{Q}^{\sigma},\sigma}^{\dagger},
\end{align}
with coefficients
\begin{align}
    \stackanchor{u_{\vek,\sigma}}{(v_{\vek,\sigma})} &=\left[\frac{1}{2}\left(1\stackanchor{+}{(-)}\frac{\frac{1}{2}(\epsilon_{\vek+\ve{Q}^{\sigma},\sigma}+\epsilon_{-\vek+\ve{Q}^{\sigma},\sigma})}{\sqrt{\big[\frac{1}{2}(\epsilon_{\vek+\ve{Q}^{\sigma},\sigma}+\epsilon_{-\vek+\ve{Q}^{\sigma},\sigma})\big]^2 +\big|\Delta_{\vek}\big|^2}}\right)\right]^{1/2}.
    \label{eq:u_and_v}
\end{align}
Using the above relation, we also express the coupling terms in Eqs.~\eqref{eq:Hc_2c1a} and~\eqref{eq:Hc_2c2a} in terms of the new fermion operators.
Note that while the coefficients $u_{\vek,\sigma}$ and $v_{\vek,\sigma}$ remain invariant under inversion of $\vek$, a finite supercurrent ($\veQs\neq0$) breaks the momentum inversion symmetry of the eigenenergies $E_{\vek,\sigma}$.

\prlsection{To the first order in perturbation theory}
We first investigate what happens if we only take into account first order terms in perturbation theory. This is done by evaluating the expectation values with respect to the new fermion operators, using that
\begin{align}
    \big<\gamma_{\vek,\sigma}^{\dagger}\gamma_{\vek',\sigma'}\big>_{\gamma} &=f_{\text{FD}}(E_{\vek-\veQs,\sigma})\delta_{\vek,\vek'}\delta_{\sigma,\sigma'},\label{eq:exp_value_finite}\\
    \big<\gamma_{\vek,\sigma}^{\dagger}\gamma_{\vek',\sigma'}^{\dagger}\big>_{\gamma}&=\big<\gamma_{\vek,\sigma}\gamma_{\vek',\sigma'}\big>_{\gamma} =0 \label{eq:exp_value_zero},
\end{align}
where $f_{\text{FD}}(E_{\vek,\sigma})$ is the Fermi-Dirac (FD) distribution. Neglecting terms that are constant in the boson operators, we find that
\begin{align}
&\left<H\right>_{\gamma}=\sum_{\ve{q}}(\omega_{\veq}+\omega_{\lambda})a_{\veq}^{\dagger}a_{\veq},\\
&\omega_{\lambda} = \frac{\lambda}{\sqrt{2NS}}\sum_{\vek,\sigma}\sigma\{(u_{\vek,\sigma}^2 -v_{\vek,\sigma}^2 ) f_{\text{FD}}(E_{\vek,\sigma})
+v_{\vek,\sigma}^2 \}.
\end{align}
Since the coupling only results in a $\veq$ independent renormalization of the magnon gap, there is no magnon spin current generated. The renormalization of the magnon gap is caused by the spin-splitting of the quasi-particle energy spectrum and exists even in the absence of supercurrents. Since the spin-splitting originates from the adjacent FI, there is an excess population of spin-up electrons, and the magnon gap is always increased. A further renormalization occurs in the presence of supercurrents due to the tilting of the quasi-particle energy spectrum.

\prlsection{To the second order in perturbation theory}
Next, we want to perform a SW transformation to obtain an effective Hamiltonian to the second order in perturbation theory \cite{bravyi_ap_11}. We define a transformed Hamiltonian $H_{\text{eff}}=e^{iS}He^{-iS}$, and use the Baker–Campbell-Hausdorff formula to expand it. By requiring that
\begin{equation}
    H_{\text{c}}^{2c1a}+H_{\text{c}}^{2c2a}=i[H_{\text{FI}}+H_{\text{S}},S] ,
\label{eq:requirement}
\end{equation}
we project out first order terms that do not commute with $H_{\text{FI}}+H_{\text{S}}$. Keeping terms up to the second order, the effective Hamiltonian can be written as
\begin{equation}
    H_{\text{eff}}=H_{\text{FI}}+H_{\text{S}} +\frac{i}{2}[S, H_{\text{c}}^{2c1a}+H_{\text{c}}^{2c2a}],
\end{equation}
where $H_{\text{c}}^{2c}$ is absorbed into $H_{\text{S}}$.
In order to determine $S$, we make an ansatz that it consists of two types of terms $S^{2c1a}$ and $S^{2c2a}$ that are of the same form as $H_{\text{c}}^{2c1a}$ and $H_{\text{c}}^{2c2a}$, respectively. The coefficients are determined by solving Eq.~\eqref{eq:requirement}.
By disregarding terms above second order in the fluctuations $a_{\veq}^{(\dagger)}$, and also neglecting all terms related to renormalization of the superconductivity caused by magnons (terms with four fermion operators), we are left with only one nonzero commutator $[S^{2c1a},H_{\text{c}}^{2c1a}]$. 
After computing the commutator, evaluating the expectation value of the Hamiltonian with respect to the fermion operators, and diagonalizing with respect to the boson operators, we end up with an effective Hamiltonian of the form
\begin{align}
\left<H_{\text{eff}}\right>_{\gamma}=&\sum_{\veq}\Omega_{\veq}\alpha_{\veq}^{\dagger}\alpha_{\veq},\\
\Omega_{\veq}=&\phantom{+}\Omega^+_{\veq+(\veQu-\veQd)}+\Omega_{-\veq-(\veQu-\veQd)}^- ,\label{eq:Omega}\\
\Omega_{\veq}^{\pm} =&\pm\frac{1}{2}\left(M^{(11)}_{\veq}-M^{(11)}_{-\veq}\right)\notag\\
&+\sqrt{\left[\frac{1}{2}\left(M^{(11)}_{\veq}+M^{(11)}_{-\veq}\right)\right]^2 -\left(M^{(12)}_{\veq}\right)^2 },\label{eq:Omega_pm}\\
M^{(11)}_{\veq}=&\frac{1}{2}\omega_{\veq-(\veQu-\veQd)}+\sum_{\vek}A_{\vek,\veq}^{(11)},\\
M^{(12)}_{\veq}=&\sum_{\vek}A_{\vek,\veq}^{(12)} .
\end{align}
The coefficients $A^{(11)}_{\vek,\veq}$ and $A^{(12)}_{\vek,\veq}$ are given in the SM \cite{SM}.
The old magnon operators are related to the new ones by
\begin{align}
    a_{\veq-(\veQu-\veQd)}&=x_{\veq} \alpha_{\veq-(\veQu-\veQd)}-w_{\veq}\alpha^{\dagger}_{-\veq-(\veQu-\veQd)},
\end{align}
with coefficients
\begin{align}
    \stackanchor{x_{\veq}}{(w_{\veq})} =& \left\{\frac{1}{2}\left[\stackanchor{+}{(-)}1+\frac{\frac{1}{2}\left(M_{\veq}^{(11)}+M_{-\veq}^{(11)}\right)}{\sqrt{\left[\frac{1}{2}\left(M_{\veq}^{(11)}+M_{-\veq}^{(11)}\right)\right]^2 -\left(M^{(12)}_{\veq}\right)^2}}\right]\right\}^{1/2}.
\end{align}
We can now evaluate expectation values of the new operators according to
\begin{align}
\big<\alpha_{\veq}^{\dagger}\alpha_{\veq'}\big>_{\alpha}&=f_{\text{BE}}\left(\Omega_{\veq}  \right)\delta_{\veq,\veq'},\\
\big<\alpha_{\veq}^{\dagger}\alpha_{\veq'}^{\dagger}\big>_{\alpha}&=\big<\alpha_{\veq}\alpha_{\veq'}\big>_{\alpha}=0,
\end{align}
where $f_{\text{BE}}(\Omega_{\veq})$ is the Bose-Einstein (BE) distribution.

Finally, we define the magnon spin current density \cite{rezende_jap_19,okuma_prl_17} polarized along $\ve{z}$ traveling in the $\ve{\delta}$ direction,
\begin{align}
    j^z_{\ve{\delta}} &= \frac{1}{N}\sum_{\veq}v_{\veq}^{\ve{\delta}} \big<S_{\veq}^z \big>_{\alpha},\label{eq:magnon_current}\hspace{0.5em}
    v^{\ve{\delta}}_{\veq} = \big(\Omega_{\veq+\ve{\delta}}-\Omega_{\veq-\ve{\delta}}\big)/2|\ve{\delta}| ,\\
    \big<S_{\veq}^z \big>_{\alpha}
    &=-[|x_{\veq+(\veQu-\veQd)}|^2 +|w_{-\veq-(\veQu-\veQd)}|^2]f_{\text{BE}}(\Omega_{\veq}).\label{eq:spin_of_momentum_mode}
\end{align}
Above, $v_{\veq}^{\ve{\delta}}$ is the velocity associated with a momentum mode $\veq$, and $\big<S_{\veq}^z \big>_{\alpha}$ is
the spin associated with the new magnon operators for a momentum mode $\veq$.

\prlsection{The induced magnon spin current}
Before presenting the results for the magnon spin current, we argue that Curie's principle allows for charge and spin supercurrents to induce magnon spin currents traveling along the same axis. Curie's principle states that an effect should obey the symmetries of the cause. The symmetry of our system is that of a square lattice with a magnetization along~$\ve{z}$. The $p$-wave order parameter does not impose any further restrictions on the symmetry. Since the presence of a charge or spin supercurrent in the plane perpendicular to the $z$ axis breaks both the $C^4$ rotational symmetry around the $z$ axis and the mirror symmetry in the $xy$ plane, these leave the system without any symmetry restrictions and allow for resulting magnon spin currents in any direction. A spin or charge supercurrent along the $z$ axis however only allows for a magnon spin current along the same axis. Since the expression for the magnon spin current density (Eq.~\eqref{eq:magnon_current}) is the same regardless of whether the supercurrent is traveling parallel or perpendicular to the magnetization, the magnon spin current has to abide by the strictest restriction. Thus, charge and spin supercurrents can only induce magnon spin currents along the same axis. 
In the following, we only present the results for two coupled S and FI 1D chains, since these are qualitatively the same as for two coupled square lattice layers.

\begin{figure}[tb]
    \centering
    \includegraphics[width=\columnwidth]{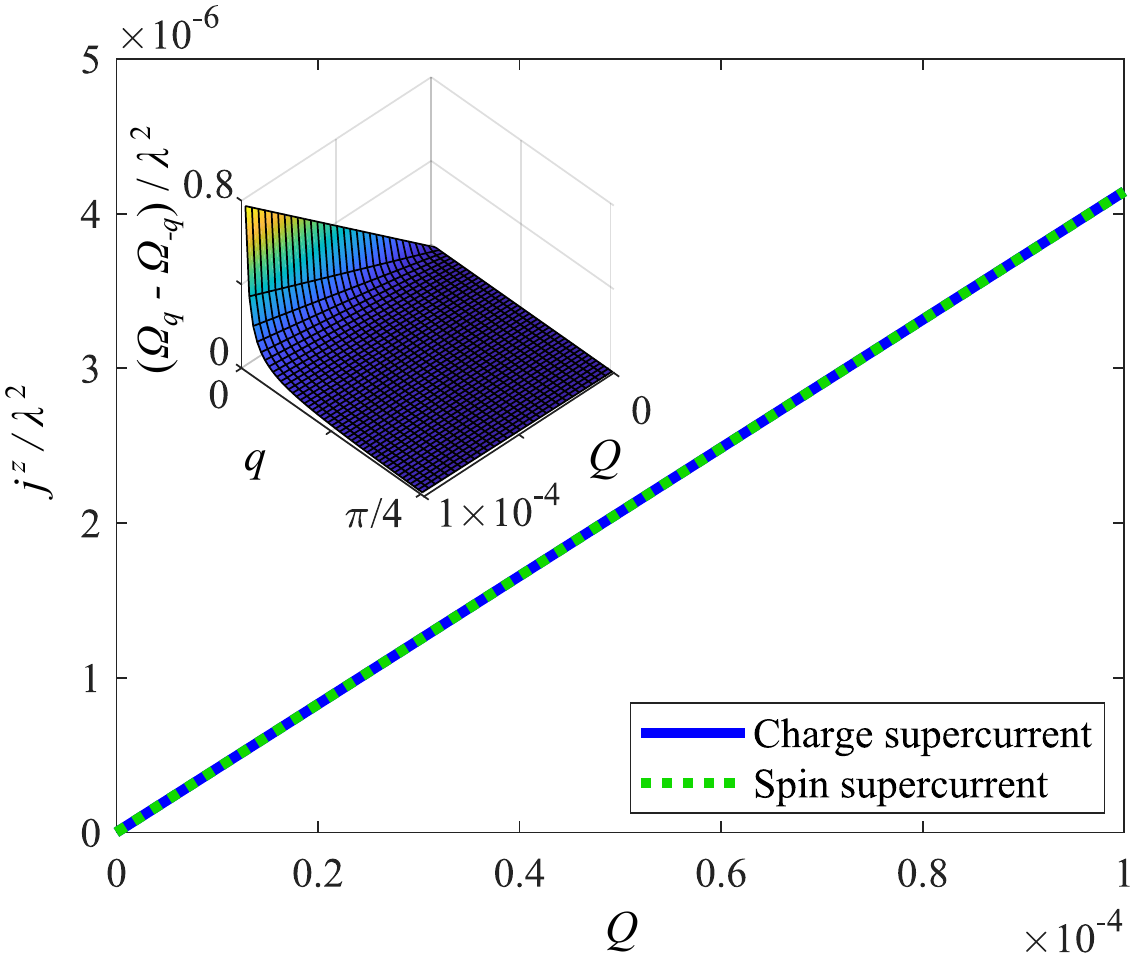}
    \caption{Both a charge ($Q^{\uparrow}=Q^{\downarrow}=Q$) and a spin ($Q^{\uparrow}=-Q^{\downarrow}=Q$) supercurrent induce the same finite magnon spin current density $j^z$ in an adjacent ferromagnetic insulator. This is because spin-up Cooper pairs dominate in inducing the magnon spin current. The supercurrents induce the magnon spin current through an asymmetry in the eigenenergy spectrum $\Omega_{q}$ of the new boson operators with respect to inversion of the momentum $q$ (inset). The results are robust to changes in the parameters, and for this particular figure we have chosen $\mu=-0.5$, $\lambda=10^{-4}$, $f_{\uparrow}=f_{\downarrow}=10^{-4}$, $S=1/2$, $J=10^{-2}$, $K=10^{-3}$ for a 1D chain of $N = 365$ lattice sites at temperature $T=0.001$. For a square lattice with a $p_{x}$- or $p_y$-wave symmetry, the results are qualitatively the same as in 1D. A supercurrent of spin-up Cooper pairs induces a current of spin-down magnons in the opposite direction, while transverse currents are not allowed. All energies, as well as the magnon spin current density, are scaled by the hopping parameter $t$. The temperature is scaled by $t/k_{\text{B}}$, where $k_{\text{B}}$ is the Boltzmann constant. The momenta $q$ and $Q$ are both scaled by $\hbar/a$, where $\hbar$ is the reduced Planck constant and $a$ is the lattice constant.}
    \label{fig:results}
\end{figure}

In Fig.~\ref{fig:results}, we show that both a charge and a spin supercurrent induce the same finite magnon spin current in the adjacent FI. A finite center of mass momentum ($Q^{\sigma}\neq0$) of the spin-$\sigma$ Cooper pairs causes a symmetry breaking in the eigenenergies $E_{k,\sigma}$ of the superconductor (Eq.~\eqref{eq:E}) with respect to inversion of the momentum. The momentum inversion symmetry is broken irrespective of the relative directions of $Q^{\uparrow}$ and $Q^{\downarrow}$. The asymmetry is transferred to the eigenenergy spectrum $\Omega_{q}$ of the new boson operators (Eqs.~\eqref{eq:Omega} and~\eqref{eq:Omega_pm}) via the coupling (Eq.~\eqref{eq:Hc_2c1a}).
To second order, the coupling transfers momentum between quasi-particles and magnons.
The eigenenergies $\Omega_q$ enter into the expression for the magnon spin current through the velocity of each momentum mode $v_{q}$ and the Bose-Einstein distribution function $f_{\text{BE}}(\Omega_q )$ in Eqs.~\eqref{eq:magnon_current}-\eqref{eq:spin_of_momentum_mode}. Supercurrents of spin-$\sigma$ Cooper pairs are thus able to introduce a momentum inversion symmetry breaking that causes a finite magnon spin current.
We illustrate this lack of symmetry by plotting $\Omega_{q}-\Omega_{-q}$ for $q>0$ in the inset of Fig.~\ref{fig:results}. The asymmetry in the magnon energy spectrum disappears for large momenta $q$, because the renormalization of the quasi-particle energy spectrum is also small for large momenta (see Eq.~\eqref{eq:E}).

The contribution from the spin-down Cooper pairs in inducing the magnon spin current is negligible compared to the contribution from the spin-up Cooper pairs. This can be seen from the results for the charge and spin supercurrents in Fig.~\ref{fig:results}. A charge supercurrent corresponds to spin-up and spin-down Cooper pairs traveling in the same direction, while for a spin supercurrent the spin-down Cooper pairs instead travel in the opposite direction.
The fact that inverting the propagation direction of the spin-down Cooper pairs has no visible effect on the magnon spin current density implies that it is mostly spin-up Cooper pairs that contribute in inducing the magnon spin current. 
The fact that spin-up and spin-down Cooper pairs do not contribute equally follows from the coupling between spins in the superconductor and ferromagnetic insulator, where the FI breaks the up-down symmetry by having a majority spin polarization along $\ve{z}$. We observe that this asymmetry must be introduced through  Eq.~\eqref{eq:Hc_2c1a}, since neglecting the spin-splitting term (Eq.~\eqref{eq:Hc_2c}) only slightly alters the results. 

\prlsection{Outlook}
Our results provides a proof-of-principle, showing that it is possible to convert both charge and spin supercurrents into a magnon spin current. Although we presented results for a $p_x$-wave symmetry in one dimension, we expect similar results with a different symmetry, \eg a $p_x +ip_y$-wave symmetry \cite{mackenzie_rmp_03}. In fact, from results for a square lattice, we have verified that we obtain qualitatively the same results for a $p_x$- and a $p_y$-wave superconducting order parameter when the supercurrent travels along the $x$ axis.
Other structures that have been shown to carry spin-polarized supercurrents include $s$-wave/ferromagnetic half-metal junctions \cite{eschrig_prl_03,keizer_n_06,sprungmann_prb_10,anwar_apl_12}, $s$-wave/magnetic multilayer structures \cite{khaire_prl_10,klose_prl_12,gingrich_prb_12}, and structures with spiral magnetic order \cite{houzet_prb_07,robinson_sc_10,alidoust_prb_10,halasz_prb_11}. These structures may also be relevant,
since the crucial ingredient for a broken momentum inversion symmetry is the presence of spin-polarized triplet supercurrents, and not the particular symmetry of the superconducting order parameter.
We therefore posit that together with the $p$-wave superconductor discussed in this work, supercurrents of odd-frequency triplets could also be potential candidates for inducing magnon spin currents in an adjacent ferromagnetic insulator.

\begin{acknowledgements}
This work was supported by the Research Council of
Norway through its Centres of Excellence funding scheme,
Project No. 262633 "QuSpin".
\end{acknowledgements}

\appendix

\section*{Supplemental Material}

In Appendix~\ref{sec:Hamiltonian}, we present further details on the derivation of our momentum space Hamiltonian from a lattice model. In Appendix~\ref{sec:coefficients}, we present expressions for $A^{(11)}_{\vek,\veq}$ and $A^{(12)}_{\vek,\veq}$ from the Eqs.~(20) and~(21) in the main text.

\section{The Hamiltonian}
\label{sec:Hamiltonian}

\subsection{The superconductor}

The momentum space Hamiltonian for the superconductor (Eqs.~(1) and~(2) in the main text) is derived starting from
\begin{align}
H_{\text{S}} = &-\sum_{\left<\ve{i},\ve{j}\right>,\sigma}t_{\ve{i},\ve{j}}c_{\ve{i},\sigma}^{\dagger}c_{\ve{j},\sigma}
-\sum_{\ve{i},\sigma}\mu_{\ve{i}}c_{\ve{i},\sigma}^{\dagger}c_{\ve{i},\sigma}\notag\\
&-\frac{1}{4}\sum_{\left<\ve{i},\ve{j}\right>,\sigma}U_{\ve{i},\ve{j}}n_{\ve{i},\sigma}n_{\ve{j},\sigma}.
\end{align}
The first term describes nearest-neighbor hopping between sites $\ve{i}$ and $\ve{j}$, where $t_{\ve{i},\ve{j}}$ is the hopping integral. The second term introduces a chemical potential $\mu_{\ve{i}}$ at lattice site $\ve{i}$. In the third term, $U_{\ve{i},\ve{j}}>0$ gives rise to the attractive nearest-neighbor interaction associated with superconductivity.
We assume the structure to be either a one-dimensional (1D) chain, or a square lattice. 
Above, $c_{\ve{i},\sigma}$ and $c_{\ve{i},\sigma}^{\dagger}$ are second quantization annihilation and creation operators for electrons with spin $\sigma$ at lattice site $\ve{i}$, and $n_{\ve{i},\sigma}\equiv c_{\ve{i},\sigma}^{\dagger}c_{\ve{i},\sigma}$ is the number operator.
We treat the superconducting term by a mean-field approach, where we assume that $c_{\ve{i},\sigma}c_{\ve{j},\sigma}\approx\left<c_{\ve{i},\sigma}c_{\ve{j},\sigma}\right>+\delta_{\text{S}}$ and neglect second order terms in the fluctuations $\delta_{\text{S}}$. We define the superconducting order parameter as
\be
\Delta_{\ve{i},\ve{j}}^{\sigma}\equiv U_{\ve{i},\ve{j}}\left<c_{\ve{i},\sigma}c_{\ve{j},\sigma}\right>.
\ee
We want to describe a superconductor that carries a charge or spin supercurrent. Such a supercurrent can be enforced by an externally applied phase gradient. To have a finite supercurrent in the system, we must require that the Cooper pairs have a finite center-of-mass velocity. This is satisfied if we write the superconducting order parameter on the form
\be
\Delta_{\ve{i},\ve{j}}^{\sigma}=\Delta_{\ve{i}-\ve{j}}^{\sigma}e^{i\ve{Q}^{\sigma}\cdot(\ve{i}+\ve{j})},
\ee
describing a propagation of the center-of-mass of a spin-$\sigma$ Cooper pair along the $\veQs$ direction \cite{takashima_prb_17,lin_sc_14}. The assumption that the phase gradient $2\veQs$ is homogeneous follows from current conservation. The superconducting order parameter can be written on this form as long as the center-of-mass velocity is small enough that the amplitude $\Delta_{\ve{i}-\ve{j}}^{\sigma}$ is independent of the center-of-mass coordinate. 
To arrive at the momentum space Hamiltonian in Eqs.~(1) and~(2) in the main text, we assume all parameters to be constant throughout the whole material, perform the Fourier transform (FT)
\be
c_{\ve{i},\sigma}=\frac{1}{\sqrt{N}}\sum_{\ve{k}}c_{\ve{k},\sigma}e^{i\vek\cdot\ve{i}},\label{eq:FTc}
\ee
and use the relation
\be 
\frac{1}{N}\sum_{\ve{i}}e^{i(\vek+\vek')\cdot\ve{i}}=\delta_{\vek,\vek'}.
\ee
Here, $N$ is the total number of lattice sites.  

Note that although we choose to set the strength of the order parameter $f_{\sigma}^{\ve{\delta}}$ in Eq.~(2) in the main text to a constant value, it can in principle be solved for self-consistently. 
This parameter is real and has the value $f_{\sigma}^{\ve{\delta}}=\Imag[\Delta^{\ve{\delta}}_{\sigma}]$,
where
\begin{align}
\Delta^{\ve{\delta}}_{\sigma}=-\frac{U}{N}\sum_{\vek}\big<c_{\vek+\veQs,\sigma}c_{-\vek+\veQs,\sigma}\big>e^{i\vek\cdot\ve{\delta}}.
\end{align}

\subsection{The ferromagnetic insulator}

The momentum space Hamiltonian for the ferromagnetic insulator (Eq.~(3) in the main text), we start from the Heisenberg Hamiltonian
\be
H_{\text{FI}}=-\sum_{\left<\ve{i},\ve{j}\right>}J_{\ve{i},\ve{j}}\ve{S}_{\ve{i}}\cdot\ve{S}_{\ve{j}}-\sum_{\left<\ve{i},\ve{j}\right>}K_{\ve{i},\ve{j}}S_{\ve{i}}^z S_{\ve{j}}^z .
\ee
The first term describes the ferromagnetic coupling between neighboring spins for $J_{\ve{i},\ve{j}}>0$. The second term introduces an anisotropy favoring alignment of the spins along the $z$ direction for $K_{\ve{i},\ve{j}}>0$. Above, $\ve{S}_{\ve{i}}$ is the total spin and $S_{\ve{i}}^z$ the spin polarized along the $z$ axis at lattice site $\ve{i}$. The spin is assumed to mainly be directed along the $z$ axis with weak spin fluctuations only. We can therefore apply a Holstein-Primakoff (HP) transformation to the second order in the magnon annihilation and creation operators $a_{\ve{i}}$ and $a_{\ve{i}}^{\dagger}$,
\begin{align}
    S_{\ve{i}}^+ &= \sqrt{2S}a_{\ve {i}},\label{eq:S+}\\
    S_{\ve{i}}^- &= \sqrt{2S}a_{\ve {i}}^{\dagger},\label{eq:S-}\\
    S_{\ve{i}}^z &= S-n_{\ve {i}}.\label{eq:Sz}
\end{align}
Above, $S$ is the magnitude of the spins in the ferromagnetic insulator, $S_{\ve{i}}^{\pm}=S_{\ve{i}}^x\pm iS_{\ve{i}}^y $, and $n_{\ve{i}}=a^{\dagger}_{\ve{i}}a_{\ve{i}}$ is the number operator. By assuming that $J_{\ve{i},\ve{j}}$ and $K_{\ve{i}}$ are constant, and applying a similar FT as for the fermion operators,
\begin{align}
    a_{\ve{i}}=\frac{1}{\sqrt{N}}\sum_{\veq}a_{\veq}e^{i\veq\cdot\ve{i}},\label{eq:FTa}
\end{align}
we arrive at the momentum space Hamiltonian for the ferromagnetic insulator given in Eq.~(3) in the main text.

\subsection{The coupling}

In order to describe a coupling between spins at the interface between the superconductor and the ferromagnetic insulator, we introduce
\begin{align}
H_{\text{c}}&=-\sum_{\ve{i}}\Lambda\ve{s}_{\ve{i}}\cdot\ve{S}_{\ve{i}},
\end{align}
where $\Lambda>0$ is the coupling strength and
\begin{align}
\ve{s}_{\ve{i}}&=\frac{1}{2}\sum_{\sigma,\sigma'}c_{\ve{i},\sigma}^{\dagger}\ve{\sigma}_{\sigma,\sigma'}c_{\ve{i},\sigma'},
\end{align}
is the electron spin and $\ve{\sigma}$ is the vector of Pauli matrices.
We apply the HP transformation in Eqs.~\eqref{eq:S+}-\eqref{eq:Sz} to the spin $\ve{S}_{\ve{i}}$ in the ferromagnetic insulator. Performing the dot product between the two spin operators, we obtain
\begin{align}
    H_{\text{c}} =& -\Lambda\sum_{\ve{i}}\sigma n_{\ve{i},\sigma}(S - a_{\ve{i}}^{\dagger}a_{\ve{i}})\notag\\
    &-\Lambda \sqrt{2S} \sum_{\ve{i}}(c_{\ve{i},\uparrow}^{\dagger}c_{\ve{i},\downarrow}a_{\ve{i}}^{\dagger}+c_{\ve{i},\downarrow}^{\dagger}c_{\ve{i},\uparrow}a_{\ve{i}}).
\end{align}
The first sum comes from the $z$ component of the spins and gives rise to spin-splitting in the quasi-particle energy spectrum and the renormalization of the magnon gap. The second sum comes from the $x$ and $y$ components and gives rise to magnon spin currents in the ferromagnetic insulator if supercurrents are present in the superconductor.
We apply the FT in Eq.~\eqref{eq:FTc} to the fermion operators and the FT in Eq.~\eqref{eq:FTa} to the magnon operators in order to arrive at the momentum space Hamiltonian for the coupling given in Eqs.~(4)-(6) in the main text. In these equations we have renamed the coupling terms to $\lambda=\Lambda\sqrt{S/2N}$ for simplicity of notation.

\begin{widetext}
\section{Expressions for the coefficients}
\label{sec:coefficients}

The coefficients $A^{(11)}_{\vek,\veq}$ and $A^{(12)}_{\vek,\veq}$ in Eqs.~(20) and~(21) in the main text are given by
\begin{align}
    A_{\vek,\veq}^{(11)}&=\frac{\lambda^2}{2}\bigg\{(u_{\vek,\uparrow})^2 (u_{\vek+\veq,\downarrow})^2 \frac{[f_{\text{FD}}(E_{\vek,\uparrow})-f_{\text{FD}}(E_{\vek+\veq,\downarrow})]}{(E_{\vek,\uparrow}-E_{\vek+\veq,\downarrow})+\omega_{\veq-(\veQu-\veQd)}}
    +(v_{\vek,\uparrow})^2 (u_{\vek+\veq,\downarrow})^2 \frac{[(1-f_{\text{FD}}(E_{-\vek,\uparrow}))-f_{\text{FD}}(E_{\vek+\veq,\downarrow})]}{-(E_{-\vek,\uparrow}+E_{\vek+\veq,\downarrow})+\omega_{\veq-(\veQu-\veQd)}}\notag\\
    &+(u_{\vek,\uparrow})^2 (v_{\vek+\veq,\downarrow})^2 \frac{[f_{\text{FD}}(E_{\vek,\uparrow})-(1-f_{\text{FD}}(E_{-\vek-\veq,\downarrow}))]}{(E_{\vek,\uparrow}+E_{-\vek-\veq,\downarrow})+\omega_{\veq-(\veQu-\veQd)}}
    +(v_{\vek,\uparrow})^2 (v_{\vek+\veq,\downarrow})^2 \frac{[(1-f_{\text{FD}}(E_{-\vek,\uparrow}))-(1-f_{\text{FD}}(E_{-\vek-\veq,\downarrow}))]}{-(E_{-\vek,\uparrow}-E_{-\vek-\veq,\downarrow})+\omega_{\veq-(\veQu-\veQd)}}\bigg\},\\
    A^{(12)}_{\vek,\veq}&=\frac{\lambda^2}{2}u_{\vek+\veq,\uparrow}u_{\vek,\downarrow}\text{sgn}(\Delta_{\vek+\veq}^{\uparrow})v_{\vek+\veq,\uparrow}\text{sgn}(\Delta_{\vek}^{\downarrow})v_{\vek,\downarrow}
    \bigg\{\frac{[f_{\text{FD}}(E_{\vek+\veq,\uparrow})-f_{\text{FD}}(E_{\vek,\downarrow})]}{-(E_{\vek+\veq,\uparrow}-E_{\vek,\downarrow})+\omega_{\veq-(\veQu-\veQd)}}
    -\frac{[f_{\text{FD}}(E_{\vek+\veq,\uparrow})-(1-f_{\text{FD}}(E_{-\vek,\downarrow}))]}{-(E_{\vek+\veq,\uparrow}+E_{-\vek,\downarrow})+\omega_{\veq-(\veQu-\veQd)}}\notag\\
    &-\frac{[(1-f_{\text{FD}}(E_{-\vek-\veq,\uparrow})-f_{\text{FD}}(E_{\vek,\downarrow})]}{(E_{-\vek-\veq,\uparrow}+E_{\vek,\downarrow})+\omega_{\veq-(\veQu-\veQd)}}
    +\frac{[(1-f_{\text{FD}}(E_{-\vek-\veq,\uparrow})-(1-f_{\text{FD}}(E_{-\vek,\downarrow}))]}{(E_{-\vek-\veq,\uparrow}-E_{-\vek,\downarrow})+\omega_{\veq-(\veQu-\veQd)}}\bigg\}.
\end{align}
\end{widetext}


%

\end{document}